\documentclass[10pt,journal,compsoc]{IEEEtran}

\usepackage{algorithm} 
\usepackage{algorithmic}
\usepackage{booktabs}
\usepackage{clrscode3e}
\usepackage{hyperref}
\usepackage{multirow}
\usepackage{graphicx}
\usepackage{subfigure}
\usepackage{textcomp}
\usepackage{xcolor}
\usepackage{acronym} 
\usepackage{amsfonts}
\usepackage{amssymb}
\usepackage{amsthm,amsmath}
\usepackage{mathrsfs}
\usepackage{indentfirst}
\usepackage{cases}
\usepackage{tabularx}
\usepackage{threeparttable}

\ifCLASSOPTIONcompsoc
  \usepackage[nocompress]{cite}
\else
  \usepackage{cite}
\fi
\ifCLASSINFOpdf
\else
\fi
\acrodef{NCSR}{Non-overlapping Cross-domain Sequential Recommendation}
\acrodef{PLCR}{Prompt Learning-based Cross-domain Recommender}
\acrodef{SASRec}{Self-Attention based Sequential Recommendation model}
\acrodef{SR}{Sequential Recommendation}
\acrodef{CSR}{Cross-domain Sequential Recommendation}
\acrodef{CR}{Cross-domain Recommendation}
\acrodef{NCR}{Non-overlapping Cross-domain Recommendation}

\begin{document}
\title{Automated Prompting for Non-overlapping Cross-domain Sequential Recommendation}
%
%
%
%
\author{Lei~Guo,
        Chunxiao~Wang,
        Xinhua~Wang,
        Lei~Zhu,
        Hongzhi~Yin,~\IEEEmembership{Senior Member,~IEEE}
\IEEEcompsocitemizethanks{\IEEEcompsocthanksitem L. Guo, C. Wang, X. Wang, and L. Zhu are with the School of Computer Science and Technology, Shandong Normal University, Jinan 250014, China.\protect\\
E-mail: leiguo.cs@gmail.com, wangchunxiao0217@163.com, wangxinhua@sdnu.edu.cn, leizhu0608@gmail.com

\IEEEcompsocthanksitem H. Yin is with the School of Information Technology \& Electric Engineering, The University of Queensland, St Lucia, QLD 4072, Australia.\protect\\
E-mail: h.yin1@uq.edu.au
}
\thanks{Manuscript received XX XX, 2023; revised XX XX, 2023.\\
(Corresponding author: Hongzhi Yin.)}}

%
%


\markboth{IEEE TRANSACTIONS ON KNOWLEDGE AND DATA ENGINEERING,~Vol.~XX, No.~X, X~2023}
{Shell \MakeLowercase{\textit{et al.}}: Bare Demo of IEEEtran.cls for Computer Society Journals}
%



\IEEEtitleabstractindextext{%
\begin{abstract}
\acf{CR} has been extensively studied in recent years to alleviate the data sparsity issue in recommender systems by utilizing different domain information.
In this work, we focus on the more general \acf{NCSR} scenario. \ac{NCSR} is challenging because there are no overlapped entities (e.g., users and items) between domains, and there is only users' implicit feedback and no content information.
Previous \ac{CR} methods cannot solve \ac{NCSR} well, since (1) they either need extra content to align domains or need explicit domain alignment constraints to reduce the domain discrepancy from domain-invariant features, (2) they pay more attention to users' explicit feedback (i.e., users' rating data) and cannot well capture their sequential interaction patterns, (3) they usually do a single-target cross-domain recommendation task and seldom investigate the dual-target ones.
Considering the above challenges, we propose \acf{PLCR}, an automated prompting-based recommendation framework for the \ac{NCSR} task.
Specifically, to address the challenge (1), \ac{PLCR} resorts to learning domain-invariant and domain-specific representations via its prompt learning component, where the domain alignment constraint is discarded.
For challenges (2) and (3), \ac{PLCR} introduces a pre-trained sequence encoder to learn users' sequential interaction patterns, and conducts a dual-learning target with a separation constraint to enhance recommendations in both domains.
Our empirical study on two sub-collections of Amazon demonstrates the advance of \ac{PLCR} compared with some related SOTA methods.
\end{abstract}

}

\maketitle

\IEEEdisplaynontitleabstractindextext

%
\IEEEpeerreviewmaketitle


%
%
%
%

 

\IEEEraisesectionheading{\section{Introduction}\label{sec:introduction}}

\IEEEPARstart{D}{}ue to the effectiveness of the cross-domain information in alleviating the sparsity issue in recommender systems~\cite{yin2019social,chen_try_2020,wang2020next}, the \acf{CR} task that aims at improving cross-domain recommendations is gaining immense attention.
Moreover, because of the solid temporal relationship between user interactions~\cite{yin2016spatio,yin_2015_MM}, the \acf{CSR} task that can model users' sequential interests becomes increasingly attractive. One typical assumption of these methods is that there are totally or partially overlapped entities between domains~\cite{guo2021gcn,guo2022reinforcement,jiang2019heterogeneous,liu2021leveraging}, by which they can make more effortless knowledge transfer across domains.

In this paper, we study \ac{CSR} in a particularly challenging scenario, \acf{NCSR}, where users and items are entirely disconnected between domains, and only users' positive interactions can be observed (no extra content information is available). We consider the non-overlapping characteristic because it is a common phenomenon
in practice.
For example, the items are entirely disjointed in movie-watching and book-reading services, and because of the business privacy policies, it is impractical to have overlapped users from different platforms/systems.
The non-existence of overlapped entities and extra auxiliary information makes it harder to transfer helpful information across domains, since there is no direct connection between them.

\begin{figure}
    \centering    \includegraphics[width=9cm]{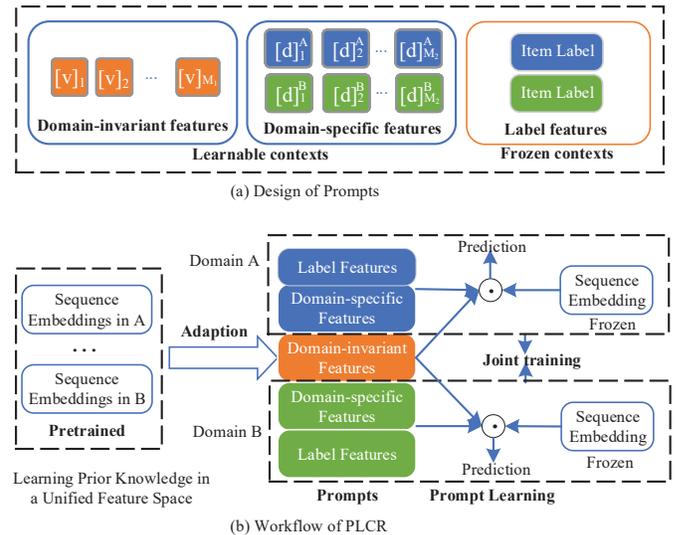}
    \caption{The design of our prompts and the workflow of \ac{PLCR}, which first models the prior knowledge under a unified feature space by a joint pre-training on both domains, and then adapts the prior knowledge to specific domains via prompt learning. The learned prompts are shared within the source/target domains between sequences, representing the common interaction patterns between users and items. These can also be interpreted as users' cluster-level interests.
    We achieve knowledge transfer without domain alignment by mapping prior knowledge in a unified feature space, and sharing the domain-invariant contexts of prompts across domains.}
    \label{fig:prompts}
\end{figure}

Previous works on \ac{NCR} mainly focus on distilling domain information from rating data, and their performance on users' sequential data is largely unexplored.
In addition, current solutions on \ac{NCR}
mainly focus on learning domain-invariant features across domains, with which the domain information can be shared between domains to enhance recommendations in each domain~\cite{liu2022collaborative,li2009can,wang2019recsys,zhang2022cross,gao2019cross}.
One common practice of such methods is based on the adversarial training~\cite{wang2019recsys,perera2019cngan,zhang2021deep},
which tends to narrow the gap between source and target domains by confusing the domain discriminators.
However, reducing the discrepancy by conducting domain alignment could result in the missing of some semantic information (i.e., the user preferences/intents that a sequence contains), due to the entangled nature of semantic and domain information. This situation may get worse when the data distributions have complex manifold structures.
To repair this, recent methods advocate preserving the semantic information to maintain the recommender's prediction capability~\cite{Kanagawa2019Cross}. But as the objectives of domain alignment and holding semantic features could be adversarial, these methods often suffer from a subtle trade-off between these two objectives.
An alternative solution to such an issue could be learning disentangled semantic and domain representations~\cite{cao_disenCDR2022,Choi_review2022}, where the domain alignment constraint is discarded. But existing studies either ignore the non-overlapped characteristic~\cite{cao_disenCDR2022} or need extra content to bridge domains~\cite{Choi_review2022}.
Furthermore, existing studies on \ac{NCR} mainly focus on single-target tasks that aim to transfer domain knowledge from the dense source domain to the sparse target domain. In contrast, the dual-target tasks that tend to enhance recommendations in both domains are seldom investigated.

\begin{figure}
    \centering    \includegraphics[width=9cm]{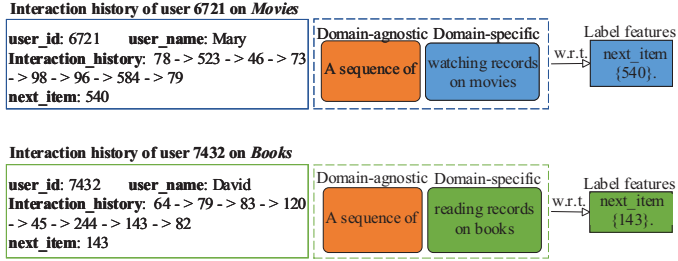}
    \caption{Examples of our designed prompts. Domain-agnostic contexts represent the general features independent of domains (shared by both domains), and domain-specific contexts denote the features related to specific domains (shared by the users within the same domain).}
    \label{fig:exampe_prompts}
\end{figure}

To address the above challenges, in this work, we target \ac{NCSR} and propose a \acf{PLCR} via leveraging the power of automated prompt engineering paradigm~\cite{ge2022domain}. The workflow of \ac{PLCR} is shown in Fig.~\ref{fig:prompts}, which firstly embeds the prior knowledge within domains into pre-trained models, and then dynamically adapts them to specific domains via sharing the prompts across domains to conduct recommendations on disconnected domains.
Compared with existing works, our approach exploits the pre-trained sequential models with prompt optimization and adaption, avoiding the domain alignment constraint and the need of extra content with only a few parameters to be optimized.
Specifically, we design each prompt into three components (Fig.~\ref{fig:prompts} shows our design on prompts): domain-independent, domain-specific, and label context, where the domain-independent context denotes domain-invariant/semantic features and is shared among all sequences with the same label. The domain-specific context represents the common information across domains. The item label refers to features of the ground truth item, which will be frozen during optimization.
Examples of our designed prompts are shown in Fig. \ref{fig:exampe_prompts}.

Then, to conduct a dual-target cross-domain recommendation, and ensure the learned semantic and domain representations are disentangled in such a continuous feature space (we do not exploit the discrete prompts due to they are by no means guaranteed to be optimal for downstream tasks),  
we leverage a separation constraint for optimization (as shown in Fig.~\ref{fig:overview}).
Concretely, we design two optimization objectives on the basis of the compound prompts by sharing the domain-independent features between domains, and then make a duel-target cross-domain recommendation with compound learning objectives.
In our learning paradigm, the domain-agnostic features are shared between domains, and expected to help reconstruct users' common interests in both domains, while the domain-specific features are expected to model domain-independent aspects that are disentangled from domain-specific features.

We summarize our main contributions as follows:
\begin{itemize}
    \item We study \ac{CSR} in a more general yet challenging scenario, i.e., \ac{NCSR}. To overcome the limitations of previous studies, we explore \ac{NCSR} by proposing an automated prompting paradigm to model the non-overlapping characteristic.  
    \item We design an automated prompt engineering paradigm to adapt the pre-trained sequential model to specific domains. With shared prompt contexts, we can transfer information across disjoint domains.
    \item We develop a separation constraint to facilitate the learned domain-invariant and domain-specific features capturing different aspects of domains to support the dual-target cross-domain recommendation.
    \item We evaluate our method by conducting extensive experiments on two sub-collections of Amazon, and the experimental results show the advance of our proposal compared with the SOTA baselines.
\end{itemize}
\section{Related Work}
\noindent In this section, we consider cross-domain recommendation with overlapped/non-overlapped entities, 
 and prompt learning-based recommendation as our related works.

\subsection{Cross-domain Recommendation with Overlapped Entities}
\noindent Existing studies on transferring domain knowledge across domains have been focused on developing domain alignment methods via overlapped users, items, or part of them. 
Among these methods, few of them tend to assume both users and items are totally overlapped between domains, and enhance the recommendations by leveraging the users' common interests on the same item set~\cite{pan2010transfer,pan2013transfer,zhu2021unified,liu2015non}.
For example, Pan et al.~\cite{pan2013transfer} target to reduce the data sparsity issue in CF domains, and solve it by proposing a collective factorization method, where a  shared latent space is collectively constructed.
To enhance cross-domain recommendations, Zhu et al.~\cite{zhu2021unified} 
first learn users' and items' embeddings via separating heterogeneous graphs and then devise an element-wise network to join the embeddings of shared entities between domains.
As it is too ideal to assume that both domains have identical users and items, recent works have focused on studying the scenarios in that only users are overlapped~\cite{liu2021leveraging,li2020ddtcdr,liu2020cross,yuan2019darec,zhu2019dtcdr,zhu2020graphical,cui2020herograph,guo2021gcn,guo2022reinforcement,sun2021parallel,guo2022time,cao2022contrastive,ma2022mixed}.
For instance, Guo et al.~\cite{guo2021gcn} investigate the cross-domain sequential recommendation task with overlapped users by proposing a network-based solution, where they enhance the representations of users by combining them with users' preferences in the other domain with a graph convolutional network.
Li et al.~\cite{li2020ddtcdr} propose a dual-transfer cross-domain recommender to conduct cross-domain recommendations by investigating the user interests in both domains.
Considering the privacy issue of aligning users, Gao et al.~\cite{gao2019cross} only view the overlapped items as bridges to connect domains, and distill helpful information from transferred item embeddings by leveraging the power of neural networks in learning high-dimensional representations.

However, as the above methods need overlapped entities to connect domains, their application may be limited when entities are entirely non-overlapping.
To remedy this, we target the non-overlapping cross-domain recommendation task and aim to solve the knowledge transfer issue by proposing a prompt learning-based method.

\subsection{Cross-domain Recommendation with Non-overlapped Entities}
\noindent Compared with overlapped or partially overlapped methods, the non-overlapping cross-domain recommendation task is more challenging, since the direct connections between domains are missing.
According to different recommendation targets, recent works in this category can be divided into single-target, dual-target, or multi-target, and one-to-many cross-domain recommendation methods.
Single-target cross-domain recommendation task tends to exploit useful knowledge from source domain to enhance the recommendations in the target domain, aiming at alleviating the cold-start and sparsity issues within the target domain~\cite{cremonesi2014cross,wang2019recsys,liu2022collaborative,li2009can,zhang2022cross,manotumruksa2019cross,perera2019cngan}.
For example,
Liu et al.~\cite{liu2022collaborative} study the review-based \ac{NCR} task by proposing an attribution alignment-based collaborative
filtering method. But their work needs extra review and attribution information, which may not be available in real-world scenarios, to align domains, limiting its application to real systems.
Dual-target or multi-target cross-domain recommendation task tends to conduct a two-way or multi-way cross-domain recommendation that two or multiple domains can benefit from knowledge exchanges~\cite{zhang2021deep,li2022recguru ,li2009transfer}. For example, Zhang et al.~\cite{zhang2021deep} match the feature spaces of both users and items between domains by learning the shared encoders with a domain discriminator under a dual-adversarial learning framework.
Li et al.~\cite{li2009transfer} investigate the multi-target cross-domain recommendation task by proposing a rating-matrix generative model, where an implicit cluster-level rating matrix is established to find the relatedness across multiple domains.
Studies on one-to-many cross-domain recommendation tasks mainly focus on transferring helpful domain information from the dense source domain to new sparse target domains~\cite{krishnan2020transfer,li2009can,cremonesi2014cross,yu2020semi}. For instance, Krishnan et al.~\cite{krishnan2020transfer} study the non-overlapping cross-domain recommendation task via sharing the domain-independent component across the dense and sparse domains.

But existing works on disjoint domains mainly focus on the single-target cross-domain recommendation task, and only a few consider enhancing the recommendation in both domains.
Moreover, the above works mainly explore the domain knowledge from the rating data, and their performance on the sequential data is still unproven.
Furthermore, none of the existing works study knowledge transfer by leveraging the prompt learning technique, which is the main task of this work.

\subsection{Prompt Learning-based Recommendation}
\noindent Prompting is a kind of prepending instructions to 
the inputs and pre-trained language models~\cite{petroni2019language} that aims at prompting the downstream tasks.
Due to its effectiveness in tuning pre-trained models, recent studies have  successfully applied it to NLP tasks~\cite{jiang2020can,lester2021power,li2021prefix,liu2021pre,petroni2019language,shin2020autoprompt}, and several of them have also investigated the application of it to the recommendation tasks~\cite{li2022personalized,geng2022recommendation,cui2022m6,wu2022personalized,wu2022selective,xin2022rethinking,liu2023pre,kuo2022zero}.
For example, Li et al.~\cite{li2022personalized} introduce prompt learning to the explainable recommendation task. They tend to enhance the explanation generation performance of recommender systems by bridging the gap between continuous prompts and the
pre-trained model.
To give users the flexibility in choosing the sensitive attributes, Wu et al.~\cite{wu2022selective} devise a prompt-based fairness-aware recommender by introducing the prompting technique from NLP to the recommender systems.
Geng et al.~\cite{geng2022recommendation} study the knowledge transfer issue between tasks by treating recommendation as language processing, and propose a unified and flexible text-to-text paradigm named P5 for recommendations.
But their work needs the same dataset for training and testing, which is different from ours, as we focus on learning models for totally non-overlapped datasets.

However, existing studies leveraging prompting techniques in improving recommendations are still in their infancy, and none considers the information transferring issue across disjoint domains.
\section{Methodologies\label{sec:methodologies}}
In this section, we first detail the definition of the \ac{NCSR} task, and then show how we implement \ac{PLCR} to conduct cross-domain recommendations based on prompt learning.

\subsection{Preliminaries}
\noindent Suppose we have two domains, A and B, and we have access to obtain users' sequential behaviours in both domains. The target of \ac{NCSR} is to recommend the next item a user will interact with in source or target domains based on her historical interactions. But different from traditional \ac{CSR}, \ac{NCSR} pays more attention to the non-overlapping scenario, where users and items are totally disjoint between domains.
\ac{NCSR} is a more challenging task due to the missing of the bridging entities, which can be directly leveraged for domain alignment.
To meet this challenge, we bring a new take on \ac{NCSR} via the prompt learning technique. 
In our setting, a prompt is a kind of prepending instructions to the pre-trained sequence and item encoders. They are designed to prompt these pre-trained models to guide the recommendations in both domains.
We tend to automatically find optimal prompts to provide accurate guidance rather than exploiting the manually designed ones.

More formally, let $\mathcal{V}_A= \{ A_1, A_2, ..., A_m, ..., A_M \}$ be the item set in domain A, and $\mathcal{V}_B=\{B_1, B_2, ..., B_n, ..., B_N\}$ be the item set in domain B (suppose the items in A and B do not overlap).
Then, from users' interaction logs on these items, we can achieve their sequential behaviors in domains A and B, respectively denoted by $(A_1, A_2, ..., A_m, ...)$ and $(B_1, B_2, ..., B_n, ...)$, where $A_m\in \mathcal{V}_A$ and $B_n \in \mathcal{V}_B$ are the interacted items in A and B. Due to the non-overlapping characteristic of domains, the user set and items are totally disjoint between domains. That is, we cannot obtain a user's sequential behaviours in both domains to directly leverage her interactions in one domain to enhance her interest modeling in the other.
Then, the target of conducting \ac{NCSR} can be defined as building a mapping function $f(\mathcal{x})$ that predicts a probability distribution over the item set $\mathcal{V}_A$ (or $\mathcal{V}_B$) by exploring the non-overlapping cross-domain sequential data, aiming at enhancing the recommendations in both domains.

\begin{figure*}
    \centering
    \includegraphics[width=17cm]{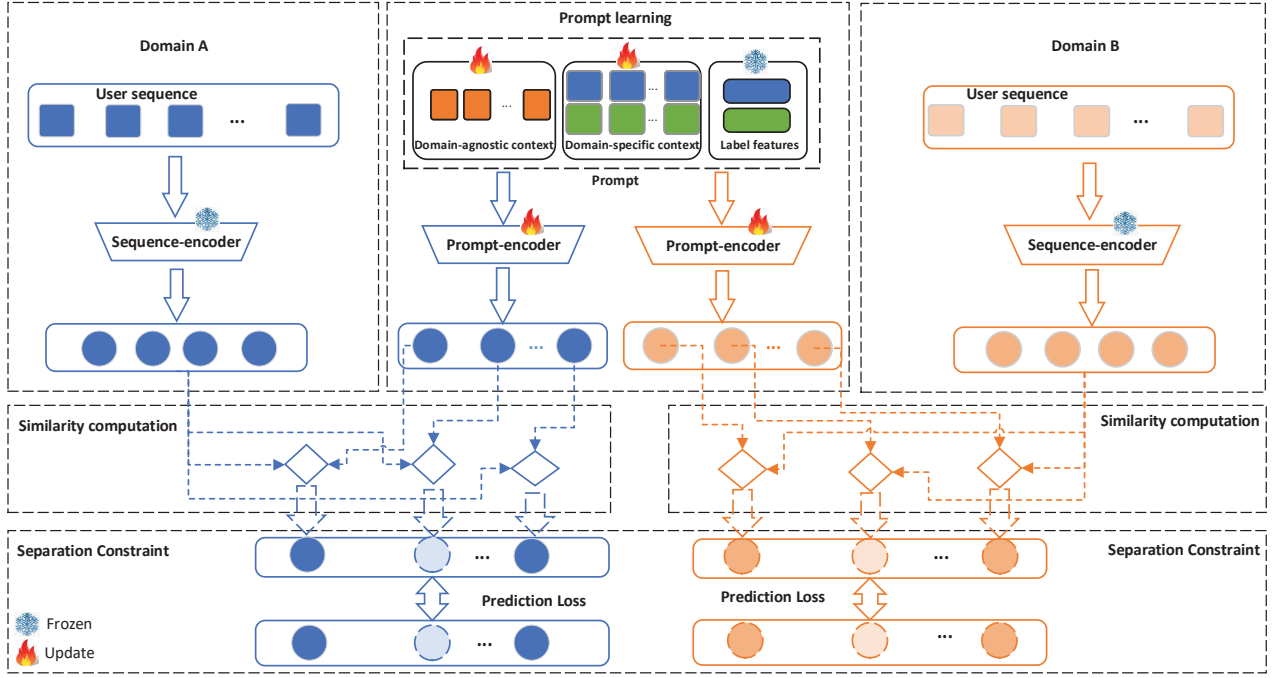}
    \caption{The system architecture of \ac{PLCR}, which consists of the pre-trained sequence and item encoders, prompt optimization module, and separation constraint component. The sequence and item encoders are pre-trained in a unified feature space. Only the prompt-encoders and prompts will be updated during prompt learning, while the pre-trained models are kept fixed.}
    \label{fig:overview}
\end{figure*}

\subsection{Overview of \ac{PLCR}}
\noindent In this work, we devise \ac{PLCR} to address the domain adaption issue in \ac{NCSR} by taking the power of prompt learning.

\textbf{Motivation.}
Due to the lack of common entities across domains, aligning two domains to make cross-domain recommendations is infeasible.
One typical line of previous methods in aligning two domains tends to impose adversarial training constraints. Still, they may result in the distortion of the domain-agnostic feature structure within domains, which can be seen as a kind of semantic-related feature unrelated to any specific domain.
Though studies have been conducted to preserve it, they still suffer from a subtle trade-off between the objectives of semantic information preservation and domain alignment.
To remedy this, we propose to learn disentangled domain-agnostic and domain-specific representations via devising a prompt learning paradigm, where the domain alignment is discarded.
The main task of this work is to embed domain information into pre-trained models, and then dynamically adapt them by learning shared prompts to enhance the cross-domain recommendations.

The overall system architecture of \ac{PLCR} is presented in Fig.~\ref{fig:overview}, which is mainly composed of three components: pre-trained session-item encoder, prompt learning, and separation constraint.
1) Session-item encoders are the pre-trained sequence and item encoding models that our method is built upon. In particular, we introduce the Transformer blocks in \cite{kang2018self} to build the encoders for sessions and items in both domains in a unified feature space. That is, we pre-train the encoders by the data in two domains simultaneously. Moreover, all the pre-trained encoders are kept fixed when conducting prompt optimization (the details of pre-training them are shown in Section \ref{section:pretraining}).
2) The prompt learning component targets transferring domain knowledge across domains by learning shared prompts in a continuous feature space, with which we can enhance cross-domain recommendations. To capture different aspects of domain information, we tend to exploit domain-specific context to model unique domain features, and domain-invariant context to learn the common features between domains
(the details of our design can be seen in Section \ref{section:prompt_learning}).
3) To ensure the learned prompt contexts are disentangled, we devise a separation constraint by sharing the domain-invariant context across domains in a dual-target learning schema. By this, the domain-invariant and domain-specific contexts can model different aspects of domains, as they have different optimization objectives (the details can be seen in Section \ref{separation}).

\subsection{Sequence and item Encoders Pre-training \label{section:pretraining}}
\noindent We adopt the self-attention blocks in \cite{kang2018self}, namely \ac{SASRec}, as the backbone of our solution, which is specially devised for sequential recommendation~\cite{vaswani2017attention} and able to draw contexts from users' all past actions.
\ac{SASRec} takes a user's interaction sequence $\mathcal{S}^A_u=(A_{u,1}, A_{u,2}, ..., A_{u,t-1})$ as input (take domain A as an example), and seeks to predict the next item at time step $t$ by exploring the previous $t-1$ items within the current sequence.
The \ac{SASRec} model comprises an embedding layer, several self-attention blocks, and a prediction layer, where the embedding layer encodes the items and their positions within the given sequence.
The self-attention block undertakes the given sequences' modeling by hierarchically extracting high-level features from all previously interacted items.
The prediction layer predicts the next item depending on the learned item and sequence embeddings.
We respectively take the self-attention block and embedding layer as the sequence-encoder and item-encoder, which will be pre-trained and helps us learn disentangled domain-specific and domain-agnostic representations.

To embed the domain information in a unified feature space, we pre-train \ac{SASRec} by training it jointly, i.e., we use the method of merging data to train the same model.
Specifically, we train \ac{SASRec} by optimizing it to align the embedding spaces learned for sequences and items, and define its learning objective as a binary cross entropy loss~\cite{kang2018self}:
\begin{align}
    -\sum_{\mathcal{S}_u\in \mathcal{S}}\sum_{t\in [1,2, ..., l_n]}\left[\text{log}(\sigma(\hat{y}_{o_t,t})) + \sum_{j\notin \mathcal{S}_u}\text{log} (1-\sigma(\hat{y}_{j,t}))\right],
\end{align}
where $\mathcal{S}_u\in \mathcal{S}$ is the user sequence in domain A or B, $l_n$ is the length of $\mathcal{S}_u$, $\hat{y}_{o_t,t}$ is the prediction probability for the expected item $o_t$ at time step $t$. $j$ is the negative item that is generated for each sequence at each time step.

As the core component (i.e., the self-attention block) of \ac{SASRec} can be accelerated through parallel training, it is one order of magnitude faster than RNN/CNN-based alternatives. That is, \ac{SASRec} is scalable to extensive training data.
However, due to the non-overlapping characteristic of NCSR and the lack of common entities that can bridge two domains, SASRec cannot be directly applied to NCSR, making us unable to adjust this pre-embedded domain knowledge across domains.

\subsection{Domain Adaption via Prompt Optimization \label{section:prompt_learning}}
\noindent One typical solution for adapting the domain knowledge from one to the other is to learn domain-invariant features, with which the recommender can further make cross-domain recommendations.
In previous studies, adversarial training is one frequently used method, which can generate domain-independent features by reducing the discrepancy between target and source domains with domain alignment.
However, as the two objectives could be adversarial, these methods often suffer from a trade-off between them.
To overcome this challenge, we resort to the prompt learning paradigm, which tends to adapt the pre-embedded domain knowledge via learning shared prompts across domains.

\subsubsection{Design of Prompts}
\noindent As shown in Fig.~\ref{fig:prompts}, our prompts are composed of three contexts, a domain-independent context, a domain-specific context, and the label features of the ground truth item.
To avoid manual prompt tuning, we model the contexts in a continuous feature space with automate prompt engineering.
During prompt optimization, only the parameters in prompts are updated, and the pre-trained models are kept fixed.
To be more specific, we use $[\boldsymbol{v}]_{m_1}, m_1 \in \{1, 2, ..., M_1\}$ to denote the token embeddings within the domain-independent context $\boldsymbol{t}_k^A$ (take domain A as an example). Its definition is shown as follows:
\begin{equation}
    \boldsymbol{t}_k^A = [\boldsymbol{v}]_1[\boldsymbol{v}]_2...[\boldsymbol{v}]_{M_1}[\boldsymbol{Item}]^A_k, 
\end{equation}
where $[\boldsymbol{v}]_{m_1}$ is the vector of the $m_1$-th context token in domain A, $M_1$ is the number of the context tokens, $\boldsymbol{t}_k^A$ is the domain-independent context for item $k$. $[\boldsymbol{Item}]_k^A$ is the embedding of item $k$, which is pre-trained with the backbone network (i.e., the SASRec model) and will be frozen during prompt learning.
The token embeddings $[\boldsymbol{v}]_{m_1}$ within the domain-independent context are shared by all the items in both domains, and aim at modeling the invariant information that is irrelevant to a specific domain, such as the users' shared interests in different domains.
The domain-independent context $\boldsymbol{t}_k^B$ in domain B can be similarly defined as:
\begin{equation}
    \boldsymbol{t}_k^B = [\boldsymbol{v}]_1[\boldsymbol{v}]_2...[\boldsymbol{v}]_{M_1}[\boldsymbol{Item}]^B_k.
\end{equation}

However, as the above token embeddings tend to be shared across domains, they cannot handle the distribution shift between domains. To repair this, we further devise a domain-specific context to model the unique features in each domain, and use $[\boldsymbol{d}]_{m_2}^A, \{m_2\in 1, 2, ..., M_2\}$ to denote the token embeddings within it. Then, the prompt in domain A can be re-defined as:
\begin{equation}
    \boldsymbol{t}_k^A = [\boldsymbol{v}]_1[\boldsymbol{v}]_2... [\boldsymbol{v}]_{M_1}[\boldsymbol{d}]_1^A [\boldsymbol{d}]_2^A... [\boldsymbol{d}]_{M_2}^A[\boldsymbol{Item}]^A_k, 
\end{equation}
where $M_2$ is the number of the domain-specific tokens, $[\boldsymbol{d}]_{m_2}^A \in \mathbb{R}^d$ is the vector of the $m_2$-th context token, and will be optimized during prompt learning. 
Both $[\boldsymbol{v}]_{m_1}$ and $[\boldsymbol{d}]_{m_2}^A$ have the exact dimensions as the embeddings in the pre-trained sequence and item encoders.
The tokens within the domain-specific context are shared among all items but specifically designed for each domain. By this, we can model domain shifts and capture users' unique interests in domains.

Other than placing the label features at the end of the prompt, we can also put them in the middle or at the front (similar definitions can also be achieved in domain B):
\begin{equation}
    \boldsymbol{t}_k^A = [\boldsymbol{v}]_1[\boldsymbol{v}]_2... [\boldsymbol{v}]_{M_1}[\boldsymbol{Item}]^A_k[\boldsymbol{d}]_1^A [\boldsymbol{d}]_2^A... [\boldsymbol{d}]_{M_2}^A,
\end{equation}
\begin{equation}
    \boldsymbol{t}_k^A =[\boldsymbol{Item}]^A_k [\boldsymbol{v}]_1[\boldsymbol{v}]_2... [\boldsymbol{v}]_{M_1}[\boldsymbol{d}]_1^A [\boldsymbol{d}]_2^A... [\boldsymbol{d}]_{M_2}^A.
\end{equation}
By this, we can increase the flexibility of prompt learning by allowing the prompt to either fill the later cells or cut off the sentence earlier. In experiments, we leave it as a hyper-parameter, and report the results in Section \ref{hyper_parameter}.

\begin{figure}
    \centering
    \includegraphics[width=7cm]{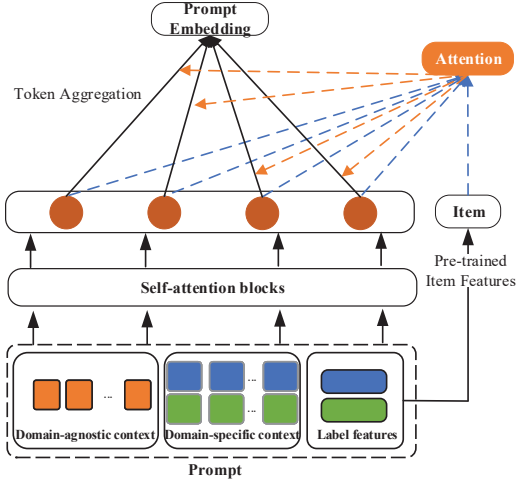}
    \caption{The design of the prompt encoder, which consists of the self-attention blocks and a token aggregation network.}
    \label{fig:prompt_encoder}
\end{figure}

\subsubsection{Prompt Encoder}
\noindent To generate high-dimensional features for prompts, we further pass them to the prompt encoder $g(\cdot)$, which is composed by several self-attention blocks and a token aggregation network to attentively aggregate the context tokens that are related to the target item (the architecture of $g(\cdot)$ is shown in Fig.~\ref{fig:prompt_encoder}).
Similar to SASRec~\cite{kang2018self}, we model the given prompt $\boldsymbol{t}_k^A$ through extracting high-level information from all the context tokens (with the head number and block number as hyper-parameters), and outputs the representations for every context word within $\boldsymbol{t}_k^A$ (take domain A as an example).

Suppose the output of the self-attention layer is denoted as:
\begin{equation}
    \boldsymbol{s}_k^A = \text{Att}(\boldsymbol{t}_k^A W^Q, \boldsymbol{t}_k^A W^K, \boldsymbol{t}_k^A W^V),
\end{equation}
where Att is the scaled dot-product attention as defined in \cite{vaswani2017attention},
$W^Q, W^K, W^V \in \mathbb{R}^{d\times d}$ are the projection matrices.
Then, to endow the extraction with nonlinearity, we further apply a point-wise two-layer feed-forward neural network to $\boldsymbol{s}_k^A$:
\begin{equation}
    \boldsymbol{f}_k^A = \text{ReLU} (\boldsymbol{s}_k^A \boldsymbol{W}^{(1)} + \boldsymbol{b}^{(1)}) \boldsymbol{W}^{(2)} + \boldsymbol{b}^{(2)},
\end{equation}
where $\boldsymbol{b}^{(1)}, \boldsymbol{b}^{(2)}$ are $d$-dimensional vectors, and $\boldsymbol{W}^{(1)}, \boldsymbol{W}^{(2)}$ are $d\times d$ matrices.
Then, we apply the layer norm operation to each block, and denote the final output of the self-attention blocks for $\boldsymbol{t}_k^A$ as $\boldsymbol{x}_k^A$.

\textbf{Token Aggregation.} To get the prompt features in the sequence level, we further aggregate the tokens within $\boldsymbol{x}_k^A$ by devising an attention mechanism with the pre-trained label features $\boldsymbol{Item}_k^A$ as the guiding signal:
\begin{equation}
    \boldsymbol{z}_k^A =Aggre(\left\{\boldsymbol{x}_{k,i}^A, \forall i \in \mathcal{C}(k)\right\})
    =\sum_{i \in \mathcal{C}(k)} \gamma_{k,i} \boldsymbol{x}_{k,i}^A,
\end{equation}
where $\mathcal{C}(k)$ is the token set within $\boldsymbol{x}_k^A$,  $\gamma_{k,i}$ implies the necessity of the $i$-th token concerning the label features, which is parameterized by a two-layer neural network:
\begin{align}
    & \gamma^*_{k,i} = \boldsymbol{w}_2^T\cdot \sigma(\boldsymbol{W}_1 \cdot [\boldsymbol{Item}^A_{k} \oplus \boldsymbol{x}^A_{k,i}] +\boldsymbol{b}_1) + \boldsymbol{b}_2
\end{align}
\begin{align}
    & \gamma_{k,i} = \frac{exp(\gamma^*_{k,i})}{\sum_{i\in \mathcal{C}(k)}exp(\gamma^*_{k,i})}
\label{eq:group_attention}
\end{align}
where $\sigma(\cdot)$ is a non-linear activation function, $\boldsymbol{W}$ and $\boldsymbol{b}$ are the weight and bias parameters, $\oplus$ is the concatenation operation, $\boldsymbol{Item}_k^A$ is the pre-trained label features of item $A_k$, and will be frozen during optimization.

\subsubsection{Prompt Optimization}
\noindent Given the training samples in both source and target domains,
we could obtain the probability that recommends item $A_k$ to the given sequence $\mathcal{S}_i^A$ (take the domain A as an example):
\begin{equation}
    p(\hat{y}^A_i = A_k |\mathcal{S}_i^A)=
    \frac{\text{exp}\left(\left <g(\boldsymbol{t}_i^A),f(\mathcal{S}_i^A)\right>\right)}{\sum_{j=1}^{N_A}exp\left(\left<g(\boldsymbol{t}_j^A),f(\mathcal{S}_i^A)\right>\right)}
\end{equation}
where $N_A$ is the number of items in domain A, $p(\hat{y}_i^A= A_k)$ indicates the matching probability of the training sample with item $A_k$, $\left<x,y\right>$ measures the correlation between the sequence embeddings and prompt representations. 

Then, we can learn prompts by optimizing a standard cross-entropy loss function on all training sequences:
\begin{equation}
    \mathcal{L}_A = -\frac{1}{|\mathcal{S}_A|} \sum_{i=1}^{|\mathcal{S}_A|} \text{log} \; p(\hat{y}^A_i = y_i^A),
\end{equation}
where $|\mathcal{S}_A|$ is the number of the training sequences in the domain A, $y_i^A$ is the ground truth of the current training sample.
Similarly, we can also get the optimization loss for the prompts in domain B:
\begin{equation}
    \mathcal{L}_B = -\frac{1}{|\mathcal{S}_B|} \sum_{i=1}^{|\mathcal{S}_B|} \text{log} \; p(\hat{y}^B_i = y_i^B),
\end{equation}
where $p(\hat{y}^B_i = y_i^B)$ denotes the probability of our method in retrieving the right item based on the given sequence.

\subsection{Separation Constraint\label{separation}}

\noindent To let the learned domain-independent and domain-specific prompt contexts can learn different aspects of the domain information, we further conduct a dual-target learning objective with the shared domain-agnostic context in both domains. The learning objective can be intuitively defined as:
\begin{equation}
    \mathcal{L} = \mathcal{L}^A + \mathcal{L}^B.
\end{equation}

However, due to the non-overlapping characteristic in \ac{NCSR}, the training samples in different domains are variant, and we cannot match them one by one, which is one of the main assumptions in traditional cross-domain methods.
As thus, instead of learning the joint loss $\mathcal{L}$, we jointly optimize $\mathcal{L}_A$ and $\mathcal{L}_B$ by devising a two-stage training method.
Specifically, we first optimize $\mathcal{L}_A$ by using the training sequences in domain A to learn domain-specific and domain-independent contexts in prompts, where the domain-independent contexts are shared across domains.
Then, besides learning domain-specific features by using the training samples in domain B, the domain-independent context within it will be further fine-tuned according to $\mathcal{L}_B$.
As $\mathcal{L}_A$ and $\mathcal{L}_B$ have different learning objectives, the domain-specific prompt contexts tend to model the features specific to each domain. In contrast, the domain-agnostic prompt context tends to learn the common features across domains.
The Stochastic Gradient Descent (SGD) algorithm is applied in our optimizing method.
\section{Experimental Setup}
\noindent In this section, we first introduce the research questions that we tend to answer in experiments, and then provide the details of the datasets, evaluation metrics, baselines, and implementations of \ac{PLCR}.

\subsection{Research Questions}
\noindent The investigated research questions in experiments are shown as follows:
\begin{itemize}
    \item[\textbf{RQ1}] How does our proposed \ac{PLCR} method perform compared with other state-of-the-art cross-domain recommenders?
    \item[\textbf{RQ2}] What are the performances of \ac{PLCR} on different domains? Can \ac{PLCR} effectively leverage the cross-domain information via the prompt tuning technique?
    \item[\textbf{RQ3}] How do the key components of \ac{PLCR}, i.e., domain-specific context, domain-independent context, the attention mechanism in prompt encoder, and the separation constraint strategy, contribute to the recommendation performance?
    \item[\textbf{RQ4}] 
    How do the hyper-parameters in \ac{PLCR} affect its performance?
\end{itemize}

\begin{table}
    \centering
    \footnotesize
     \caption{The statistics of our HAMAZON datasets.}
    \begin{tabular}{lcccccc}
    \toprule
    & \multicolumn{3}{c}{Movie-Book}
    &\multicolumn{3}{c}{Food-Kitchen} \\
    \cmidrule{1-7}
    \multirow{1}[1]{*}{}& \multicolumn{3}{c}{M-domain} & \multicolumn{3}{c}{F-domain}\\
    \#Items &\multicolumn{3}{c}{12,669} &\multicolumn{3}{c}{24,434}\\
    \#Interactions &\multicolumn{3}{c}{1,052,443} &\multicolumn{3}{c}{217,409}\\
    \#Avg.sequence length &\multicolumn{3}{c}{9.68} &\multicolumn{3}{c}{6.37}\\
    \cmidrule{1-7}
    \multirow{1}[1]{*}{}& \multicolumn{3}{c}{B-domain} & \multicolumn{3}{c}{K-domain}\\
    \#Items &\multicolumn{3}{c}{36,776} &\multicolumn{3}{c}{24,658}\\
    \#Interactions &\multicolumn{3}{c}{1,060,974} &\multicolumn{3}{c}{180,657}\\
    \#Avg.sequence length &\multicolumn{3}{c}{9.75} &\multicolumn{3}{c}{5.29}\\
    \cmidrule{1-7}
    \#Sequences &\multicolumn{3}{c}{108,770} &\multicolumn{3}{c}{34,143}\\
    \#Train-sequence & \multicolumn{3}{c}{81,578}
    & \multicolumn{3}{c}{25,608} \\
    \#Test-sequence & \multicolumn{3}{c}{10,876}
    & \multicolumn{3}{c}{5,121} \\
    \#Val-sequence & \multicolumn{3}{c}{16,316}
    & \multicolumn{3}{c}{3,414} \\
    \bottomrule
    \end{tabular}
    \label{tab:dataset_statistics}
\end{table}

\begin{table*}[ht]
\begin{center}
  \centering
  \scriptsize
   \caption{Comparison results (\%) on the Movie-Book and Food-Kitchen datasets, where M and B stand for the Movie and Book domain, F and K stand for the Food and Kitchen domain, respectively.}
    \begin{tabular}{lcccccccc|cccccccc}
    \toprule
    \multicolumn{1}{c}{\multirow{2}[4]{*}{\textbf{Methods}}} & 
    \multicolumn{4}{c}{\textbf{M-domain }} & 
    \multicolumn{4}{c|}{\textbf{B-domain }} & 
    \multicolumn{4}{c}{\textbf{F-domain}} & 
    \multicolumn{4}{c}{\textbf{K-domain}} \\
    \cmidrule{2-17}
          & \multicolumn{2}{c}{HR} & \multicolumn{2}{c}{NDCG} 
          & \multicolumn{2}{c}{HR} & \multicolumn{2}{c|}{NDCG}
          &
          \multicolumn{2}{c}{HR} & \multicolumn{2}{c}{NDCG}
          & \multicolumn{2}{c}{HR} & \multicolumn{2}{c}{NDCG} 
          \\
          \midrule
          & @10      & @20   & @10    & @20   & @10    & @20   & @10    & @20  & @10     & @20   & @10    & @20   & @10    & @20   & @10    & @20
          \\
    \midrule
    Session-POP~\cite{he2017neural}&0.55&1.60&0.20&0.47&0.45&1.25&0.17&0.37    &1.12&2.76&0.42&0.84&0.75&1.68&0.29&0.53\\
    BPR-MF~\cite{rendle2014bayesian} &0.80&1.61&0.40&0.61&0.90&1.53&0.44&0.60     &2.04&3.64&0.83&1.25&0.09&0.21&0.04&0.07\\
    GRU4REC~\cite{hidasi2015session} &2.75&3.80&1.46&1.72&0.81&1.51&0.54&0.71    &2.11&3.05&1.01&1.25&0.13&0.49&0.10&0.19\\    HRNN~\cite{quadrana2017personalizing} &2.17&3.24&1.36&1.63&1.13&2.08&0.59&0.82&2.04&2.75&0.99&1.18&0.16&0.26&0.07&0.09\\
    NARM~\cite{li2017neural} &2.16&3.37&0.93&1.23&1.18&1.80&0.51&0.67&1.59&2.55&0.66&0.90&0.64&1.10&0.27&0.38\\ Caser~\cite{tang2018personalized}&0.66&1.21&0.37&0.45&0.30&0.52&0.17&0.20&1.15&1.94&0.67&0.76&0.75&1.39&0.37&0.52\\
    SRGNN~\cite{wu2019session} &4.26&6.62&2.27&2.87&2.34&3.19&1.36&1.57&{3.35}&5.47&{1.67}&{2.19}&0.95&1.51&{0.58}&{0.80}\\ 
    SASRec~\cite{kang2018self} &{5.71} &{9.62}& {2.72}&{3.70}&{3.85}&{6.37}&{1.82}&{2.46}&{3.23} &{5.67} &{1.53} &{2.15} &{1.15} &{1.77} &{0.57} &{0.72}\\  
    CoOp~\cite{zhou2022learning} &{6.84} &{11.08}& {3.32}&{4.38}&{4.76}&{7.58}&{2.45}&{3.15}&{3.15} &{5.51} &{1.53} &{2.12} &{0.81} &{1.45} &{0.37} &{0.53}\\  
    \midrule
    $\pi$-net~\cite{ma2019pi} &4.04&6.00&2.29&2.78&1.62&2.35&1.04&1.23&2.27&3.75&1.14&1.51&0.34&0.61&0.17&0.24\\
    PSJ-net~\cite{sun2021parallel} &4.31&6.43&2.56&3.09&1.68&2.33&1.02&1.18    &2.79&4.12&1.42&1.75&0.37&0.79&0.20&0.30 \\
    DA-GCN~\cite{guo2021gcn} &1.72&2.45&0.98&1.16&1.84&2.52&1.13&1.28    &2.61&4.26&1.44&1.85&0.84&1.32&0.41&0.53\\
    SASRec (joint)~\cite{kang2018self}   &5.17&8.75&2.56&3.48&3.62&6.24&1.79&2.45&2.87&4.53&1.38&1.80&1.08&1.86&0.52&0.72\\
    \ac{PLCR} (single)&{7.09}&{11.05}&{3.52}&{4.51}&{5.23}&{8.22}&{2.59}&{3.34}&{3.54}&{5.75}&{1.76}&{2.31}&{1.15}&{1.77}&{0.59}&0.75\\
    \midrule
         \centering
         \ac{PLCR} &\textbf{8.06*}    & \textbf{12.69*}  & \textbf{4.06*}    & \textbf{5.22*}  & \textbf{7.21*}     & \textbf{11.47*}  & \textbf{3.62*}      & \textbf{4.70*} &\textbf{4.67*}&\textbf{7.43*} &\textbf{2.17*} &\textbf{2.86*} &\textbf{1.16} &\textbf{1.97*} &\textbf{0.62*} &\textbf{0.83*}   \\
    \bottomrule
    \end{tabular}%
    \begin{tablenotes}  
        \scriptsize       
        \item Significant improvements over the best baseline results are marked with * (t-test, $p<$.05).
      \end{tablenotes} 
  \label{tab:resutls}%
\end{center}
\end{table*}%

\subsection{Datasets and Evaluation Metrics}
\noindent We evaluate our proposal and baselines on two sub-collections of Amazon\footnote{https://jmcauley.ucsd.edu/data/amazon}, which is a review dataset that collects users' review behaviours on products in different domains, including userid, itemid, ratings, timestamp, and descriptions about the items, etc.
To investigate the performance of our proposed \ac{PLCR} method on \ac{NCSR}, we pick two pairs of complementary domains, i.e., ``Movie-Book" and ``Food-Kitchen", as our evaluation datasets.

To be more specific, the ``Movie-Book" dataset records the amazon users' watching and reading behaviours on the ``Movie" and ``Book" domains, respectively. 
The ``Food-Kitchen" dataset collects the purchase records of the amazon users on the ``Food" domain and ``Kitchen" domain.
As we target the non-overlapping cross-domain recommendation, we do not need to identify the users in both domains, and organize the training samples one-to-one across domains.
To filter out the invalid and noisy data, we only keep the users with more than 5 interactions, and the items whose interaction frequency is larger than 5.
We show the statistics of our resulting datasets in Table \ref{tab:dataset_statistics}.

Our experiments take the last item in each sequence as the ground truth label.
We formulate our training dataset by randomly selecting 75\% of the whole sequences, chosing 10\% as the validation set, and defining the remaining as the test set.
As the test sequences are relatively small in the Food-Kitchen dataset, we combine the validation and test set together for testing.
For evaluation, we exploit two widely used metrics, HR@$K$ and NDCG@$K$, as our evaluation metrics to investigate the performance of our method and baselines in predicting top-$K$ items for each test sequence (10 and 20 are applied). 

\subsection{Baselines\label{baselines}}
\noindent We compare \ac{PLCR} with two types of baselines, i.e., single-domain \ac{SR} methods, and \acf{CSR} methods.

1) \ac{SR} methods on single domains:
\begin{itemize}
    \item Session-POP~\cite{he2017neural}: This is a basic yet effective recommendation method that tends to predict the most popular items in each sequence as the next item to be interacted with.
    \item BPR-MF~\cite{rendle2014bayesian}: This is a commonly used matrix factorization-based recommendation method, and utilizes a pair-wise ranking objective. We apply it to \ac{SR} by encoding a sequence with the average latent factors of the items within it.
    \item GRU4REC~\cite{hidasi2015session}: This method applies GRU for the \ac{SR} task, which uses a session-parallel mini-batch training process and optimizes a ranking-based loss function.
    \item HRNN~\cite{quadrana2017personalizing}: This is an improved version of GRU4REC by further devising a hierarchical RNN structure, where the sequential relationships among sessions are further considered.
    \item NARM~\cite{li2017neural}: In this method, the authors take the attention mechanism into session modeling to capture users' main purpose from their historical behaviours.
    \item Caser~\cite{tang2018personalized}: This is a convolutional neural network-based sequential recommendation method.    
    \item SRGNN~\cite{wu2019session}: This method models the sequences as graph-structured data, and leverages a graph neural network to catch the complex transitions among items.
    \item SASRec~\cite{kang2018self}: This method proposes a self-attention based sequential model for \ac{SR}. We use this method as the backbone of our proposal, which is first pre-trained, and frozen during prompt optimization.
    \item CoOp~\cite{zhou2022learning}: This method is proposed for prompting the pre-trained vision-language models to the image recognition tasks. We adapt this method to our task by treating sequences as images, and utilizing SASRec as its backbone network.
\end{itemize}

\begin{table*}
    \caption{Results of ablation studies on the Movie-Book and Food-Kitchen datasets (\%).}
    \centering
    \scriptsize
    \begin{tabular}{lcccccccc|cccccccc}
    \toprule
    \multicolumn{1}{c}{\multirow{2}[4]{*}{\textbf{Methods}}} & 
    \multicolumn{4}{c}{\textbf{M-domain }} & 
    \multicolumn{4}{c|}{\textbf{B-domain }} & 
    \multicolumn{4}{c}{\textbf{F-domain}} & 
    \multicolumn{4}{c}{\textbf{K-domain}} \\
    \cmidrule{2-17}
          & \multicolumn{2}{c}{HR} & \multicolumn{2}{c}{NDCG}
          & \multicolumn{2}{c}{HR} & \multicolumn{2}{c|}{NDCG}
          &
          \multicolumn{2}{c}{HR} & \multicolumn{2}{c}{NDCG}
          & \multicolumn{2}{c}{HR} & \multicolumn{2}{c}{NDCG}
          \\
          \midrule
          & @10      & @20   & @10    & @20   & @10    & @20   & @10    & @20  & @10     & @20   & @10    & @20   & @10    & @20   & @10    & @20
          \\
    \midrule
    SASRec (joint)&5.17&8.75&2.56&3.48&3.62&6.24&1.79&2.45&2.87&4.53&1.38&1.80&1.08&1.86&0.52&0.72\\
    PLCR (no specific) &7.83&12.52&3.99&5.17&7.19&11.53&3.59&4.67    &4.28&6.96&1.97&2.64&1.14&1.70&0.55&0.69\\
    PLCR (no independent) &7.97&12.67&3.98&5.15&6.49&10.03&3.36&4.25    &4.64&7.21&2.10&2.75&1.18&1.64&0.58&0.70\\      
    \ac{PLCR} (no separation)&7.62&12.47&3.97&5.18&6.45&9.95&3.42&4.30&4.53&7.09&2.11&2.76&1.03&1.80&0.48&0.68\\
    \ac{PLCR} (no attention)&7.81&12.27&3.94&5.06&6.56&9.80&3.42&4.24&4.37&7.28&2.04&2.77&1.14&1.80&0.58&0.74\\
    \midrule
     \ac{PLCR} &\textbf{8.06}    & \textbf{12.69}  & \textbf{4.06}    & \textbf{5.22}  & \textbf{7.21}     & \textbf{11.47}  & \textbf{3.62}      & \textbf{4.70}   
     &\textbf{4.67}&\textbf{7.43} &\textbf{2.17} &\textbf{2.86} &\textbf{1.16} &\textbf{1.97} &\textbf{0.62} &\textbf{0.83}   \\
    \bottomrule
    \end{tabular}%
  \label{tab:ablation}%
\end{table*}%

2) \ac{CSR} methods. For the methods proposed for the overlapping \ac{CSR} task, we adapt them to \ac{NCSR} by aligning the cross-domain training samples through random replication operations. And, to simulate the non-overlapping scenario, the shuffle operation is also applied.
\begin{itemize}
    \item $\pi$-net~\cite{ma2019pi}: This is a SOTA cross-domain sequential recommendation method that exploits a parallel information-sharing network to enhance recommendations simultaneously in both domains. 
    \item PSJ-net~\cite{sun2021parallel}: This method proposes a split-join network by first splitting the mixed representations to get role-specific representations, and then joining them to get cross-domain representations.
    \item DA-GCN~\cite{guo2021gcn}: This is a graph-based cross-domain recommendation method that develops a domain-aware attentive graph neural network for \ac{CSR}.
    \item SASRec (joint)~\cite{kang2018self}: This is the cross-domain version of SASRec, which is jointly trained by the data in both domains.
    \item \ac{PLCR} (single): This is a variant of \ac{PLCR} that takes the SASRec trained on each every domain as the backbone network.
\end{itemize}

\subsection{Implementation Details}
\noindent We implement \ac{PLCR} depending on the PyTorch platform, and accelerate the training processes via a GeForce GTX TitanX GPU. We pre-train SASRec~\cite{kang2018self} on both datasets and take the self-attention blocks and the embedding layer within it as the sequence and item encoders, respectively.
We freeze the parameters in the pre-trained encoders and learn prompts with the mini-batch SGD optimizer in 200 epochs, where the learning rate is initialized as 0.0001, and the batch size is set as 128.
For all the sequences, we constrain the maximum length of sequences as 77 for the "Movie-Book" dataset, and 64 for the "Food-Kitchen" dataset.
As for the hyper-parameters, we set the block number as 2, the head number of the self-attention network as 1, and the length of the prompt contexts $M_2$ as 5. We search the length of the prompt contexts $M_1$ within [1, 5]. We leave the chosen of the label token positions as hyper-parameters, and discuss the results of picking them in Section \ref{hyper_parameter}.
For the dimension of all the latent factors, we set its value as 50.
For the hyper-parameters in
baselines, we set their values according to the papers and further fine-tune them on both datasets.

For fair comparisons, all the baselines are trained with no pre-training. Note that, although we pre-train the SASRec as our backbone network, we do not further update them during our prompt learning. Instead, they are regarded as a fixed external knowledge to train the prompts, which also does not undergo the pre-training process.
\begin{figure*}
    \centering
    \includegraphics[width=18cm]{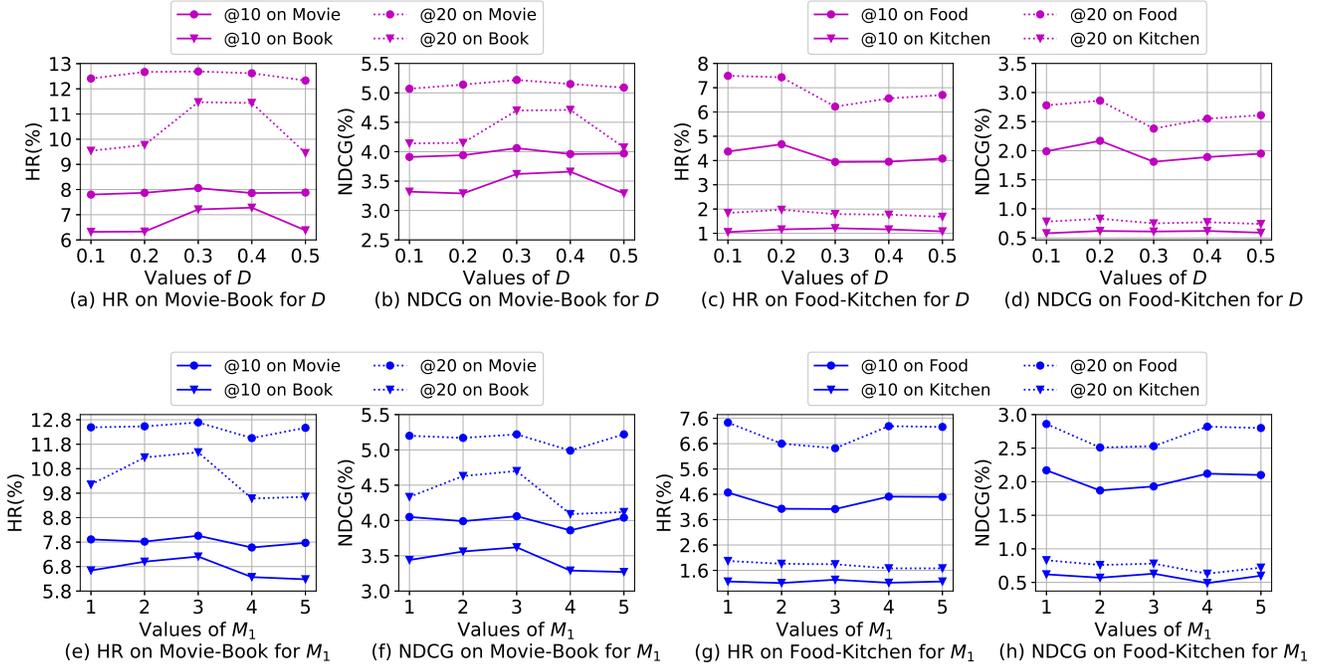}
    \caption{Impact of the hyper-parameters on Movie-Book and Food-Kitchen.}
    \label{fig:hyper_params_drop}
\end{figure*}

\section{Experimental Results (RQ1 \& RQ2)}
\noindent The experimental results are reported in Table~\ref{tab:resutls}, from which we can observe that: 
1) Our \ac{PLCR} method outperforms all the baselines on both datasets, demonstrating the superiority of leveraging the prompt learning technique in solving the \ac{NCSR} task, which can effectively transfer the domain knowledge across domains by sharing the same prompt contexts.
2) \ac{PLCR} achieves the best performance over all the baselines on both domains, showing its capability in conducting dual-target cross-domain recommendations. That is, it can effectively leverage the cross-domain information and enhance the recommendations in both domains.
3) \ac{PLCR} performs better than single-domain methods (e.g., GUR4REC, NARM, Caser, SRGNN, and SASRec), again demonstrating the benefit of utilizing the cross-domain information, which enables us to make better recommendations.
The gap between \ac{PLCR} and SASRec, indicating the effectiveness of conducting cross-domain recommendations.
4) The improvement of \ac{PLCR} over other recent SOTA cross-domain recommendation methods (i.e., $\pi$-net, PSJ-net, and DA-GCN) demonstrates the superiority of the prompt learning technique in solving the cross-domain recommendation task for non-overlapped users.
Due to the cross-domain recommendation methods (i.e., $\pi$-net, PSJ-net, and DA-GCN) are not specially devised for non-overlapped users, they cannot outperform all the single-domain methods (e.g., SRGNN, and SASRec), which may be caused by the incorrect domain matching.
5) We also notice that though we use the data in both domains to train the SASRec model (i.e., SASRec (joint)), it has worse performance than training SASRec separately, demonstrating the necessity of designing prompt learning-based cross-domain recommender, and we cannot transfer domain knowledge by simply putting the domain data together. 
6) The performance of \ac{PLCR} (single) is worse than \ac{PLCR}, indicating the importance of learning prompts in a unified feature space, which is one of the key points enabling us to transfer knowledge across domains successfully.
\ac{PLCR} (single) can achieve better results than CoOp, demonstrating the advance of our prompts in sharing domain knowledge.

\section{Experimental Analysis\label{analysis}}
\noindent In this section, we tend to answer RQ3 and RQ4 by first conducting ablation studies to evaluate the necessity of different model components, and then investigating the the key hyper-parameters' impact on \ac{PLCR}.

\subsection{Ablation Studies (RQ3)}
\noindent To explore the necessity of different model components, we compare \ac{PLCR} with four variants of it:
\begin{itemize}
    \item \ac{PLCR} (no specific): This is a variant of \ac{PLCR} that removes the domain-specific context from the prompt. This is to demonstrate the necessity of the domain-specific contexts in adapting the pre-trained models.
    \item \ac{PLCR} (no independent): This variant removes the domain-independent context from \ac{PLCR}, which tends to demonstrate the importance of domain-independent context in modeling the domain information.    
     \item \ac{PLCR} (no separation): This variant of \ac{PLCR} disables the separation constraint within it, and trains our recommendation objective separately.
     \item \ac{PLCR} (no attention): This version of \ac{PLCR} tends to evaluate the impact of the token aggregation mechanism by removing the attention network from the prompt encoder. 
\end{itemize}

\begin{figure*}
    \centering
    \includegraphics[width=18cm]{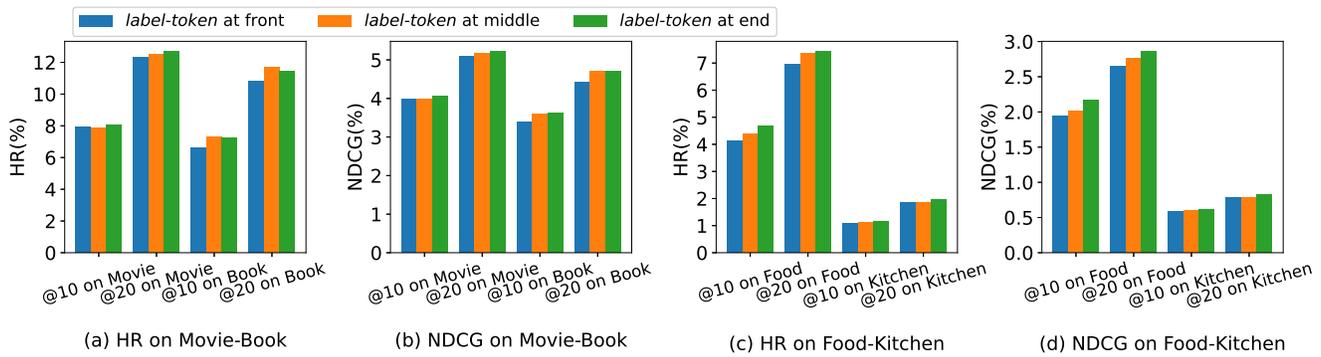}
    \caption{Impact of the location of the label token at prompts on both datasets.}
    \label{fig:hyper_params_location}
\end{figure*}

The experimental results of conducting ablation studies are shown in Table~\ref{tab:ablation}, from which we observe that:
1) \ac{PLCR} outperforms \ac{PLCR} (no specific) and \ac{PLCR} (no independent) on both datasets, indicating the effectiveness of the domain-specific and domain-independent contexts in prompting the pre-trained model to the non-overlapping cross-domain recommendation task. 
\ac{PLCR} and all its variants perform better than SASRec (joint), indicating the effectiveness of our design in model users' cross-domain preferences, and the domain information cannot be effectively utilized to enhance cross-domain recommendations by just combining data in different domains.
2) \ac{PLCR} achieves better results than \ac{PLCR} (no separation), showing the improvement of adding the separation constraint to the learning objective, which forces the domain-independent and domain-specific contexts to learn different aspects of the domain information. Moreover, this result also indicates that our prompt learning strategy can capture users' common interests across domains that can be leveraged in enhancing recommendations in both domains.
3) The gap between \ac{PLCR} and \ac{PLCR} (no attention), showing the importance of the token aggregation module in learning the high-dimensional features of the prompts.

\subsection{Hyper-parameter Analysis (RQ4)\label{hyper_parameter}}
\noindent We further explore the impact of the key hyper-parameters to \ac{PLCR} on both datasets, and report their results in Figs.~\ref{fig:hyper_params_drop} and \ref{fig:hyper_params_location}.
We explore three kinds of hyper-parameters, i.e., the dropout ratio $D$ in the prompt encoder, length of the domain-invariant contexts $M_1$, and location of the label token.
Fig.~\ref{fig:hyper_params_drop} shows the experimental results for tuning $D$ and $M_1$, from which we can observe that the performance of \ac{PLCR} changes significantly. According to our results, we set $D=0.3$ for the Movie-Book data, and $D=0.2$ for the Food-Kitchen data. A small value of $M_1$ means we will model the domain-invariant features by only a few contexts. Its impact is varied on different datasets, and the best performance is achieved on Movie-Book when $M_1=3$, and Food-Kitchen when $M_1=1$. 

To explore the impact of the label token's location, we further conduct experiments by placing the label token at the front, middle, or end of the prompts. From the results in Fig.~\ref{fig:hyper_params_location}, we can see that when we cut off the sentence earlier by the label features, the performance of \ac{PLCR} will have a slight decrease. Hence, we put the label token at the end of the prompts.
\section{Conclusions} \noindent In this work, we focus on \ac{NCSR} and solve it by proposing a \acf{PLCR}.
Specifically, to transfer the domain knowledge between domains without any overlapped entities, we leverage the power of prompt tuning, which uses the pre-trained models on both domains, and adapts the external knowledge within it to different domains by learning prompt contexts composed by domain-specific and domain-invariant features. Prompts can also be viewed as the cluster-level interests shared by all the users within the same domain, and the domain transfer is reached by sharing the domain-invariant features across domains.
To ensure the learned domain-independent and domain-specific contexts can model different aspects of the domain information, we further add a separation constraint to the learning objective via a dual-learning target.
We conduct experiments on two sub-collections of Amazon, and from the experimental results, we can demonstrate the advance of our solution in conducting cross-domain recommendations.

In this work, we mainly target developing pre-trained model-based cross-domain recommenders for the tasks with a single recommendation type, while their performance to multiple types of recommendation tasks is not explored. We will leave the cross-domain recommendation with various types as one of our future works.
\ifCLASSOPTIONcompsoc
  \section*{Acknowledgments}
\else
  \section*{Acknowledgment}
\fi
\noindent This work was supported by the Natural Science Foundation of Shandong Province (No. ZR2022MF257), and the Australian Research Council
(Nos. DP190101985, FT210100624).

\ifCLASSOPTIONcaptionsoff
  \newpage
\fi



%



\bibliographystyle{IEEEtran}
\bibliography{sample-base}

%
\vspace{-1.5cm}
\begin{IEEEbiography}
[{\includegraphics[width=1in,clip,keepaspectratio]{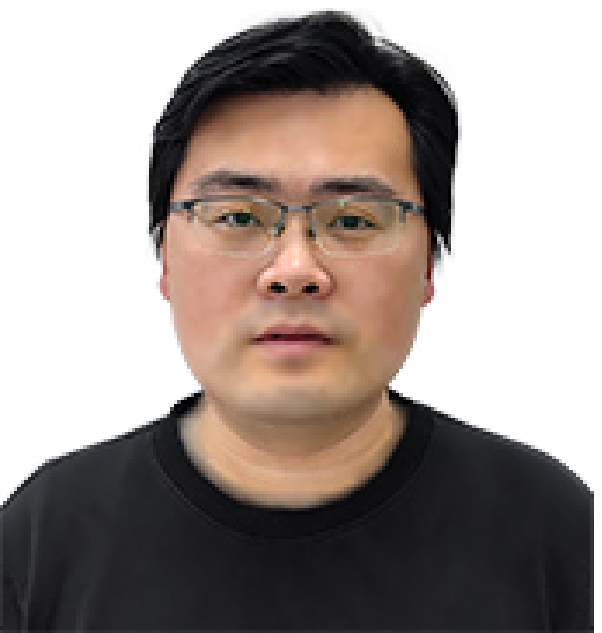}}]
{Lei Guo}
received his Ph.D. degree in computer science from Shandong University, China, in 2015. He was a Visiting Scholar with the University of Queensland from 2018 to 2019. He is currently an Associate Professor and a Master Supervisor with Shandong Normal University, China. He is a member of the Social Media Processing Committee of the Chinese Information Society. His research interests include information retrieval, social media mining, and recommender systems.
\end{IEEEbiography}
\vspace{-1cm}
\begin{IEEEbiography}
[{\includegraphics[width=1in,clip,keepaspectratio]{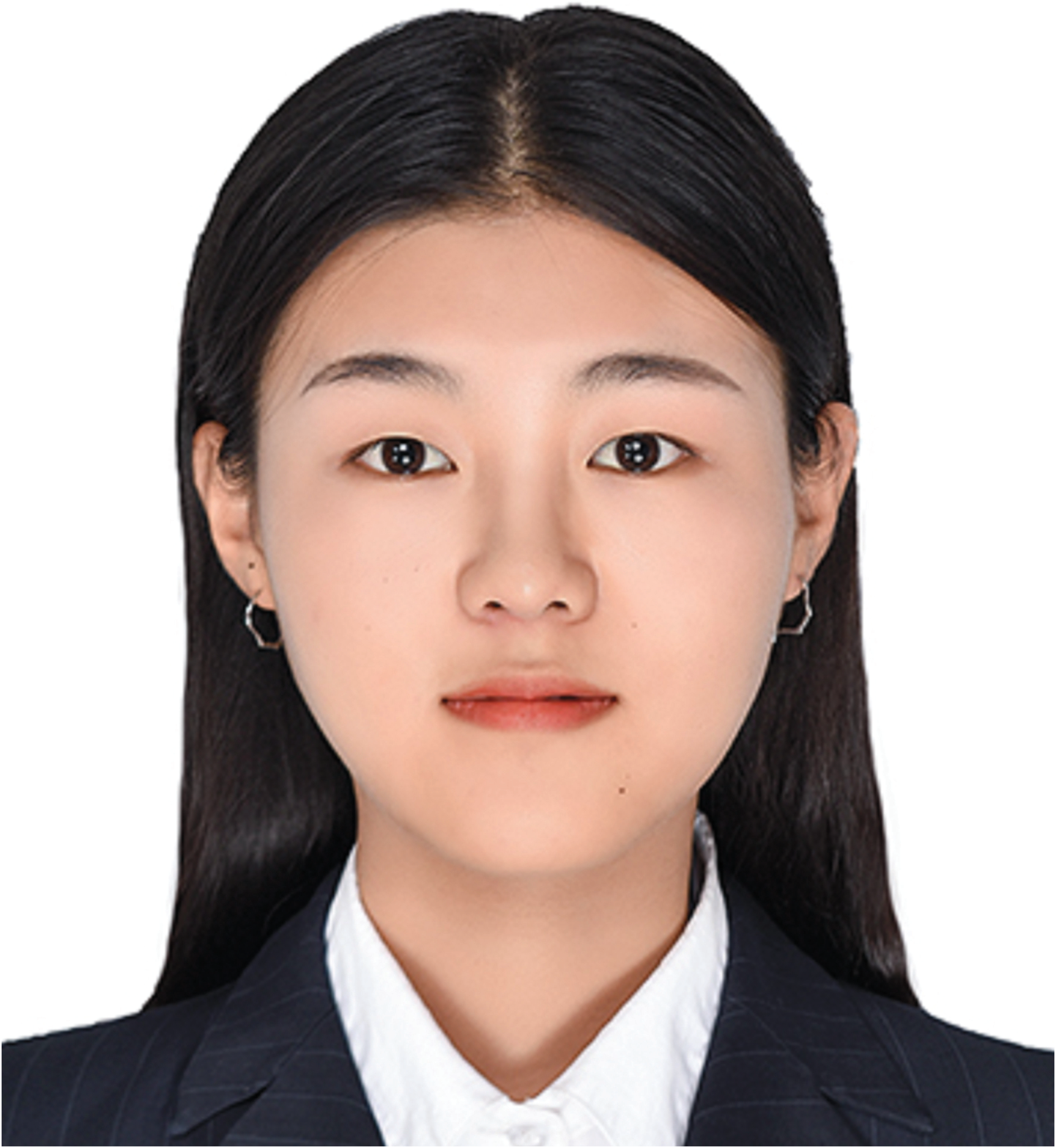}}]
{Chunxiao Wang}
is a Master candidate in computer science and technology at the School of information science and Engineering, Shandong Normal University. Her research interests include sequential recommendation and cross-domain recommendation.
\end{IEEEbiography}
\vspace{-1cm}
\begin{IEEEbiography}
[{\includegraphics[width=1in,clip,keepaspectratio]{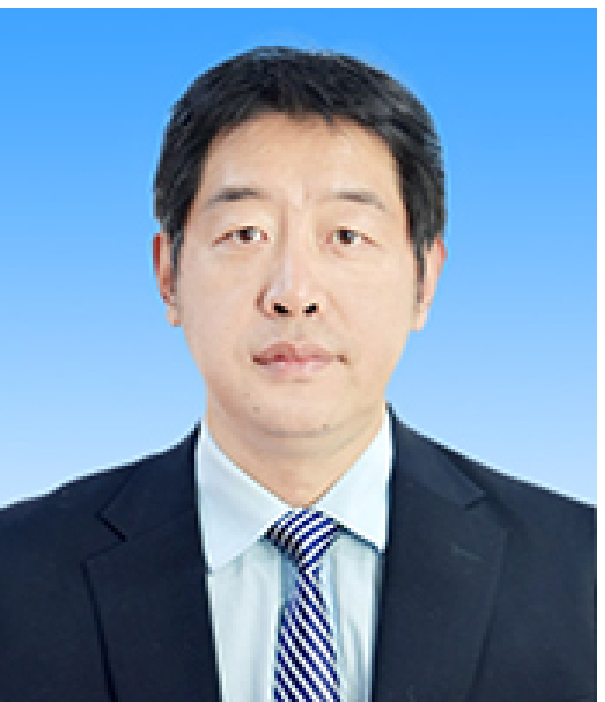}}]
{Xinhua Wang}
received the Ph.D. degree in management science and engineering from Shandong Normal University, China, in 2008. He was a Senior Visiting Scholar with Peking University, from 2008 to 2009. He is currently a Professor and a Master Supervisor with the School of Information Science and Engineering, Shandong Normal University. His research interests include distributed networks and recommender systems.
\end{IEEEbiography}
\vspace{-1cm}
\begin{IEEEbiography}
[{\includegraphics[width=1in,height=1.25in,clip,keepaspectratio]{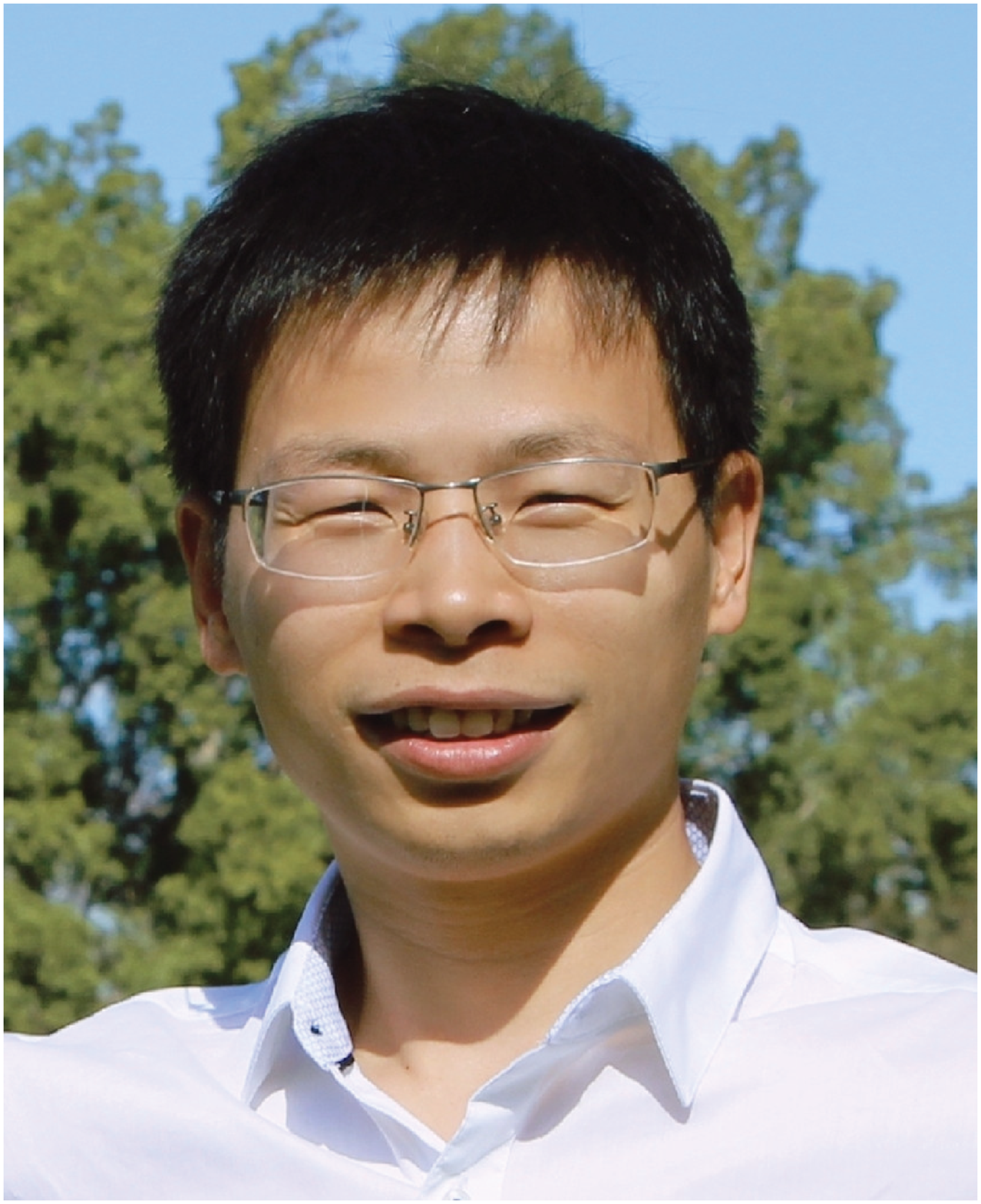}}]
{Lei Zhu} received his PH.D. degree in computer science from Huazhong University Science. He is a professor with the School of Information Science and Engineering, Shandong Normal University. 
He won ACM SIGIR 2019 Best Paper Honorable Mention Award, ADMA 2020 Best Paper Award, ChinaMM 2022 Best Student Paper Award, ACM China SIGMM Rising Star Award, Shandong Provincial Entrepreneurship Award for Returned Students, and Shandong Provincial AI Outstanding Youth Award.
His research interests are in the area of large-scale multimedia content analysis and retrieval.
\end{IEEEbiography}
\vspace{-1cm}
\begin{IEEEbiography}
[{\includegraphics[width=1in,height=1.25in,clip,keepaspectratio]{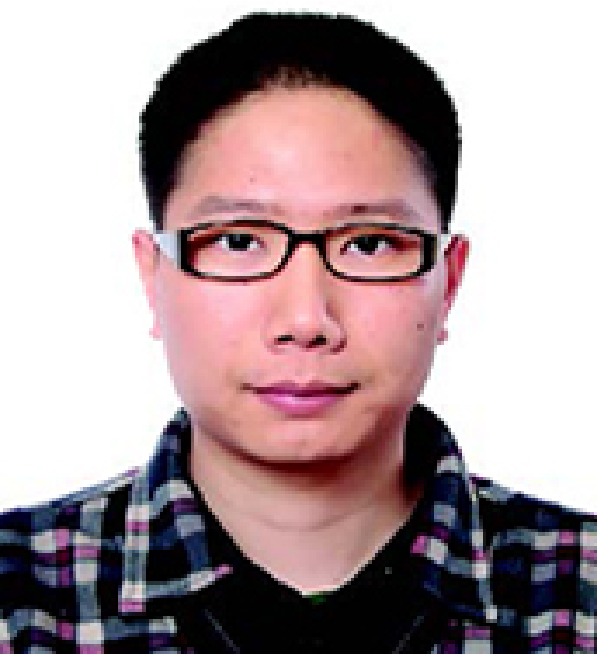}}]
{Hongzhi Yin}
received his Ph.D. degree in computer science from Peking University in 2014. He is an
Associate Professor and Future Fellow with the University
of Queensland. He received the Australian Research Council Future Fellowship and Discovery
Early-Career Researcher Award in 2016 and 2021, respectively. His research interests include recommendation system, user profiling, topic models, deep
learning, social media mining, and location-based services.
\end{IEEEbiography}





\end{document}